\documentclass[10pt]{iopart}
\usepackage{iopams,setstack,graphicx,color,bm,mathptmx,url}

\newcommand{\Gal}[1]{\mathbb{F}_{#1}}
\newcommand{\XOR}{\mbox{\textsc{xor}}}
\newcommand{\SWAP}{\mbox{\textsc{swap}}}
\newcommand{\prim}{\sigma}

\newcommand{\openone}{\leavevmode\hbox{\small1\normalsize\kern-.33em1}}

\eqnobysec

\begin{document}

\title{Graph states in phase space}
\author{A B Klimov$^1$, C Mu\~{n}oz$^{1}$ and L L S\'{a}nchez-Soto$^{2}$}

\address{$^1$ Departamento de F\'{\i}sica, Universidad de Guadalajara,
44420~Guadalajara, Jalisco, Mexico}

\address{$^2$ Departamento de \'Optica, 
Facultad de F\'{\i}sica, Universidad Complutense, 
28040~Madrid, Spain}

\date{\today}

\begin{abstract}
The phase space for a system of $n$ qubits is a discrete grid of $2^{n}
\times 2^{n}$ points, whose axes are labeled in terms of the elements of the
finite field $\Gal{2^n}$ to endow it with proper geometrical
properties. We analyze the representation of graph states in that phase
space, showing that these states can be identified with a class of
nonsingular curves. We provide an algebraic representation of the most
relevant quantum operations acting on these states and discuss the
advantages of this approach.
\end{abstract}


%\maketitle

\section{Introduction}

Graph states represent a versatile class of entangled states of
uttermost importance in quantum information. Originally proposed for
implementing measurement-based quantum
computation~\cite{Schlingemann:2001vn,Raussendorf:2001uq,Raussendorf:2003kx,van-den-Nest:2006fk}, 
they have found numerous applications in other problems such as
quantum error correction~\cite{Looi:2008ys}, secure quantum
communication~\cite {Dur:2005zr}, entanglement
purification~\cite{Dur:2003kx}, and to demonstrate fractional braiding
statistics of anyons~\cite{Han:2007vn}.  Furthermore, instances of
this family, such as Greenberger-Horne-Zeilinger (GHZ) states and
cluster states, play an essential role in fundamental tests of quantum
nonlocality~\cite{Scarani:2005ly,Guhne:2008ve}. Consequently, a great
deal of effort has been devoted to theoretically understand their
properties~\cite{Hein:2006fk,Briegel:2009cr} and to create and
manipulate them
experimentally~\cite{Lu:2007qf,Vallone:2008nx,Gao:2010bh,Gao:2010dq}.

In addition to all these applications, the graph-state formalism is a
useful abstraction that permits, in principle, a detailed (although
not exhaustive) classification of $n$-qubit
entanglement~\cite{Hein:2004fr,Danielsen:2004,Cabello:2009ly}. However,
the number of graphs grows very fast with $n$, so such a job becomes
rather involved for many qubits. Moreover, for these high-dimensional
systems, the action of nonlocal operations is difficult to interpret.

An important remark in this respect is that other quantum tasks (such
as, e.g., tomography) can be most efficiently implemented in phase
space. For $n$ qubits, the phase space is a discrete grid
of $2^{n} \times 2^{n}$
points~\cite{Wootters:1987qf,Galetti:1992ve,Gibbons:2004bh,Wootters:2004uq,Vourdas:2007dq},
whose axes are labelled by the discrete Galois field
$\Gal{2^{n}}$ in order to preserve some intuitive geometrical
properties~\cite{Lidl:1986fk}. In this way, a \textit{bona fide}
Wigner function can be
introduced~\cite{Galetti:1988uq,Vourdas:2005,Cormick:2006ty,Gross:2007nt,Bjork:2008}.
One natural question that comes to mind is whether this phase-space
picture may be of any help in reinterpreting complex issues of graph
states. Our work provides a positive answer: graph states turn out to
be nothing but discrete curves in phase space~\cite{Klimov:2009bk},
with properties closely related to other intriguing notions such as
unbiasedness~\cite{Klimov:2007fk}. In this way, quantum operations
appear as transformations on these curves, with very simple and
elegant properties.

\section{Curves in phase space}

A qubit is realized as a state in a two-dimensional Hilbert space. It
is customary to choose two normalized orthogonal states, $\{|0\rangle,
|1\rangle \}$, to serve as a computational basis. The unitary
operators
\begin{equation}
  \sigma_{z}=|0\rangle \langle 0|-|1\rangle \langle 1|\, , 
  \qquad
  \sigma_{x}=|0\rangle \langle 1|+|1\rangle \langle 0| \, ,  
  \label{sigmas}
\end{equation}
generate the Pauli group $\mathcal{P}_{1}$ of a single qubit under
matrix multiplication~\cite{Chuang:2000fk}. The elements of this group
are known as Pauli operators and provide a basis of unitary operators
on the Hilbert space.

For $n$ qubits, a compact way of labeling both states and
elements of the corresponding Pauli group $\mathcal{P}_{n}$ consists
in using the finite field $\Gal{2^{n}}$ (the reader interested
in more mathematical details is referred, to the excellent monograph
by Lidl and Niederreiter~\cite{Lidl:1986fk}). This can be considered
as a linear space spanned by an abstract basis $\{ \theta_{1}, \ldots,
\theta_{n} \}$, so that given a field element $\alpha $ (henceforth,
field elements will be denoted by Greek letters) the expansion
\begin{equation}
  \alpha = \sum_{i=1}^{n} a_{i} \,\theta_{i} \, , 
  \qquad 
  a_{i}\in \mathbb{Z}_{2} \, ,  
  \label{alpha}
\end{equation}
allows us the identification $\alpha \Leftrightarrow (a_{1}, \ldots,
a_{n})$. Moreover, the basis can be chosen to be orthonormal with
respect to the trace operation (the self-dual basis); that is,
\begin{equation}
  \tr ( \theta_{i} \, \theta_{j}) = \delta_{ij} \, ,
\end{equation}
where $\tr (\alpha )= \alpha +\alpha^{2} + \ldots + \alpha^{2^{n-1}}$,
which actually maps $\Gal{2^{n}} \mapsto \mathbb{Z}_{2}$. In
this way, we associate each qubit with a particular element of the
self-dual basis: qubit $_{i}\Leftrightarrow \theta_{i}$.

We denote by $|\alpha \rangle$, with $\alpha \in \Gal{2^n}$, an
orthonormal basis in the Hilbert space of the system. Operationally,
the elements of the basis can be labeled by powers of a primitive
element (i.e., a root of a minimal irreducible polynomial, called the
primitive polynomial), and reads $\{|0 \rangle, \, |\sigma \rangle,
\ldots, |\sigma^{2^n-1} = 1 \rangle \} $. These vectors are
eigenvectors of the operators $Z_{\beta}$ belonging to the generalized
Pauli group $\mathcal{P}_{n}$, whose generators are
\begin{equation}
  Z_{\lambda} = \sum_{\alpha} \chi ( \lambda \alpha ) \, 
  | \alpha \rangle  \langle \alpha | \, , 
  \qquad 
  X_{\lambda} = \sum_{\alpha} | \alpha + \lambda \rangle 
  \langle \alpha | \, .  
  \label{XZgf} \\
\end{equation}
Here the additive characters $\chi $ are defined as $\chi (\alpha ) =
\exp [ i \pi \tr ( \alpha ) ]$. We have then
\begin{equation}
  Z_{\alpha} X_{\beta} = \chi ( \alpha \beta ) \, X_{\beta} Z_{\alpha}\, ,
\end{equation}
which is the discrete counterpart of the Heisenberg-Weyl algebra for
continuous variables.

The operators (\ref{XZgf}) can be factorized into tensor products of
powers of single-particle Pauli operators. This factorization can be
carried out by mapping each element of $\Gal{2^{n}}$ onto an
ordered set of natural numbers according to
\begin{equation}
  Z_{\alpha} = \sigma_{z}^{a_{1}}\otimes \cdots \otimes 
  \sigma_{z}^{a_{n}} \, ,
  \qquad
  X_{\beta} = \sigma_{x}^{b_{1}}\otimes \cdots \otimes
  \sigma_{x}^{b_{n}} \, ,
\end{equation}
where $a_{i}=\tr(\alpha \theta_{i})$ and $b_{i}=\tr(\beta \theta_{i})$
are the corresponding expansion coefficients for $\alpha $ and $\beta$
in the self-dual basis.

The commutator of two monomials in $\mathcal{P}_{n}$ is
\begin{equation} 
  [ Z_{\alpha_{1}} X_{\beta_{1}}, Z_{\alpha_{2}} X_{\beta_{2}} ] = 
  [ \chi ( \alpha_{1} \beta_{2} + \beta_{1} \alpha_{2} ) - 1 ] \, 
  Z_{\alpha_{2}} X_{\beta_{2}} \, Z_{\alpha_{1}} X_{\beta_{1}} \, ,
\end{equation}
so they commute when
\begin{equation} 
\label{eq:commcond} 
\tr (\alpha_{1} \beta_{2} + \beta_{1} \alpha_{2} )=0 \, .
\end{equation}

We next recall~\cite{Wootters:1987qf,Galetti:1992ve,Gibbons:2004bh}
that the grid defining the phase space for $n$ qubits can be
appropriately labeled by the discrete points $(\alpha, \beta)$, which
are precisely the indices of the operators $Z_{\alpha}$ and
$X_{\beta}$: $\alpha$ is the ``horizontal'' axis and $\beta$ the
``vertical'' one.

From this perspective it seems reasonable to consider a (parametric)
curve in this discrete grid as a set of $2^{n} + 1$ points $(\alpha
(\kappa), \beta (\kappa))$, where the parameter $\kappa$ runs through
the field $ \Gal{2^{n}}$. The curve pass through the origin
when $( \alpha(0),\beta (0) ) =(0,0)$, and it is nonsingular (i.e.,
with no self-intersection) if all the pairs $( \alpha (\kappa ),
\beta (\kappa))$ are different.

If we represent the curve as
\begin{equation}
  \alpha (\kappa ) = \sum_{i=0}^{n-1}\alpha_{i}\,\kappa^{2^{i}} \, , 
  \qquad
  \beta (\kappa )=\sum_{i=0}^{n-1}\beta_{i}\,\kappa^{2^{i}} \, ,  
  \label{curve1}
\end{equation}
with $\alpha_{i},\beta_{i}\in \Gal{2^{n}}$, we will say that
the curve is additive commutative whenever
\begin{equation}
  \sum_{i\neq j}\tr(\alpha_{i}\beta_{j})=0\,. 
  \label{cc_g2}
\end{equation}
Due to the form in equation~(\ref{curve1}), the following property is
automatically fulfilled
\begin{equation}
  \alpha (\kappa +\kappa^{\prime}) = 
  \alpha (\kappa )+\alpha (\kappa^{\prime}) \, , 
  \qquad 
  \beta (\kappa +\kappa^{\prime})=
  \beta (\kappa )+ \beta  (\kappa^{\prime}) \, ,  
  \label{add}
\end{equation}
which means that by summing the coordinates of any two points of the
curve we obtain another point on the curve. The condition
(\ref{cc_g2}) guarantees that the monomials labeled with points of
such a curve form indeed an Abelian subgroup under multiplication
[notice the condition (\ref{eq:commcond})]; i. e.,
\begin{equation} 
  [ Z_{\alpha (\kappa)} X_{\beta (\kappa)}, 
  Z_{\alpha (\kappa^{\prime})}X_{\beta (\kappa^{\prime})}]=0 \,.
  \label{stab}
\end{equation}
Observe also (\ref{add}) implies that this group is isomorphous to
$\mathbb{Z}_{2}^{\oplus n}$.

The $2^{n}$ monomials $\{ Z_{\alpha (\kappa )}X_{\beta(\kappa )} \}$
constitute thus the whole set of stabilizer
operators~\cite{Chuang:2000fk} and $n$ of them appropriately chosen
can be considered as generators. This means that one can pick $n$
specific points that generate the whole curve by simple addition. The
simplest way is to pick the horizontal coordinates as the points
corresponding to the elements of the basis $\alpha = \{\theta
_{1},\ldots ,\theta _{n}\}$, the order of points being unessential.

The simplest form of additive commutative curves are the straight
lines
\begin{equation}
  \label{eq:sline}
  \alpha (\kappa ) = \mu \kappa \, , 
  \qquad 
  \beta ( \kappa ) = \nu \kappa \, ,
\end{equation}
which can be represented in the regular form $\alpha =0, \beta
=\lambda \alpha$. It is a well established
result~\cite{Wootters:1989fk,Wootters:2006uq} that the operators
$\{Z_{\alpha}X_{\beta =\lambda \alpha} \}$ associated to straight
lines commute for any fixed value of $\lambda \in \Gal{2^{n}}$,
while the eigenstates of the set $\{Z_{\alpha} \}$ define the standard
logical basis.

A curve is called nondegenerate (or regular) when can be represented
in the explicit form
\begin{equation}
  \beta =f(\alpha )=\sum_{i=0}^{n-1}\phi_{i} \, \alpha^{2^{i}} 
  \quad 
  \mathrm{or} 
  \quad 
  \alpha =g(\beta )=\sum_{i=0}^{n-1}\psi_{i} \, \beta^{2^{i}} \, ,
  \label{RC}
\end{equation}
with $\phi_{i},\psi_{i}\in \Gal{2^{n}}$. Such curves are
nonsingular and the commutativity condition (\ref{cc_g2}) imposes the
restrictions
\begin{equation}
  \phi_{k}=\phi_{n-k}^{2^{k}}\, , 
  \qquad 
  \psi_{k}=\psi_{n-k}^{2^{k}} \, , 
  \quad
  k=1,\ldots ,[(n-1)/2]\, ,  
  \label{cc}
\end{equation}
where $[\,]$ denotes the integer part. For even values of $n$, the
additional requirement $\phi_{n/2}=\phi_{n/2}^{2^{n/2}}$
($\psi_{n/2}=\psi_{n/2}^{2^{n/2}}$) should be satisfied.

The expansion coefficients are related to the curve equation in a very
simple form. Indeed, by rewriting the equation $\beta =f ( \alpha ) $
as
\begin{equation}
  \beta =f ( \alpha ) = 
  f \left ( \sum_{k=1}^{n} \tr ( \alpha \theta_{k} ) \theta_{k} \right) = 
  \sum_{k=1}^{n} \tr ( \alpha \theta_{k} ) f ( \theta_{k} )  = 
  \sum_{k=1}^{n} \sum_{i=0}^{n-1} \alpha^{2^{i}} 
\end{equation}
and comparing with (\ref{RC}), we immediately get that
\begin{equation}
  \phi_{i}=\sum_{k=1}^{n} \theta_{k}^{2^{i}} f( \theta_{k} ) \, .
  \label{phi_f}
\end{equation}

The degenerate curves have more involved
structure~\cite{Klimov:2009bk}. The degeneration basically means that
$\alpha $ and $\beta $ do not take some values in the field, which is
compensated by multiple appearance of other (admissible)
``coordinates'' , since the curve is nonsingular. Such curves are
characterized by the degrees of degeneration $r_{\alpha }$ and
$r_{\beta }$. This degree means the following: if $(\alpha _{j},
\beta_{j})$ is a point of a degenerate curve, for each $\alpha _{j}$
there are $2^{n-r_{\beta }}$ values of $\beta $, such that the points
$(\alpha _{j},\beta _{k})$ ($k=1,\ldots ,2^{n-r_{\beta }}$) belong to
the same curve and, conversely, for each $\beta _{j}$ there are 
$2^{n-r_{\alpha }}$ values of $\alpha $, such that the points 
$(\alpha_{k},\beta _{j})$ also belong to the same curve.

\section{Operations with curves}

We next show how the action of Clifford operations can be described in
this finite grid. For a single qubit, the simplest of such operations
are $x$- and $z$-rotations, which are represented by the unitaries
$u_{x,z}=\exp (i\pi \sigma _{x,z}/4)$. They produce the
transformations transformations
\begin{eqnarray}
  &&x \mathrm{-rotation:}\,\{\sigma _{x} 
  \mapsto 
  \sigma _{x},\sigma _{y}
  \mapsto
  \sigma _{z},\sigma _{z}
  \mapsto 
  \sigma _{y}\} \, ,  
  \nonumber \\
  && \\
  &&z \mathrm{-rotation:}\,\{\sigma _{x}
  \mapsto 
  \sigma _{y},\sigma _{y}ç
  \mapsto
  \sigma _{x},\sigma _{z}
  \mapsto 
  \sigma _{z}\}\,.  
  \nonumber
\end{eqnarray}
Multiqubit rotations can be labeled by $\xi \in \Gal{2^{n}}$, so that
in the self-dual basis the expansion $\xi =\sum_{i}\xi _{i}\theta_{i}$
indicates on which qubit the rotation is performed: if $\xi_{i}=1$,
then the $i$-th qubit is rotated. The general form of a local $x$- and
$z$-rotation is thus $U_{x,z}^{\xi}=\prod_{i=1}^{n}u_{x,z}^{\tr(\xi
  \theta _{i})}$. The indices of the monomials $Z_{\alpha }X_{\beta }$
transform then according to
\begin{eqnarray}
  &&x\mathrm{-rotation:}\left\{ 
    \begin{array}{l}
      \alpha \mapsto \alpha \,, \\ 
      \\ 
      \beta \mapsto \beta +\sum_{k}\xi _{k}
    \tr(\alpha \theta_{k})\theta _{k} \,,
    \end{array}
  \right.   \nonumber \\
  && \\
  &&z\mathrm{-rotation:}\left\{ 
    \begin{array}{l}
      \alpha \mapsto \alpha +\sum_{k}\xi _{k}\tr(\beta \theta _{k})\theta _{k}\,,  \\ 
      \\ 
      \beta \mapsto \beta \,.
    \end{array}
  \right.   \nonumber  
\label{x}
\end{eqnarray}
With local transformations it is always possible to transform any degenerate
curve into a nondegenerate one. Moreover, any curve can be cast into the
form $\beta =f(\alpha )$, where $f(\alpha )$ is an invertible function, so
that the curve equation can be also written as $\alpha =f^{-1}(\beta )$.

One- and two-qubit gates are crucial in quantum computation. As an
archetypal example, we consider the nonlocal $\XOR_{ij}$
operator (applied to the $i$th and $j$th qubits) with the following action 
\begin{eqnarray}
  \openone^{(i)} \sigma_{z}^{(j)} \mapsto
  \sigma_{z}^{(i)}\sigma_{z}^{(j)} \, ,
  \quad 
  \sigma_{z}^{(i)} \openone^{(j)} \mapsto 
  \sigma_{z}^{j} \openone^{(i)} \, , \nonumber \\
  \\ 
  \openone^{(i)} \sigma_{x}^{(j)} \mapsto 
  \openone^{(i)} \sigma_{x}^{(j)},
  \quad 
  \sigma_{x}^{(i)} \openone^{(j)} \mapsto 
  \sigma_{x}^{(i)} \sigma_{x}^{(j)} \, . \nonumber
\end{eqnarray}
In our algebraic language this can be elegantly represented by
\begin{equation}
  \label{eq:XORdef}
  \alpha \mapsto \alpha + \theta_{i} \tr ( \alpha \theta_{j}) \, , 
  \qquad
  \beta \mapsto \beta + \theta_{j} \tr ( \beta \theta_{i}) \, ,
\end{equation}
so that
\begin{equation}
  \XOR_{ij} =
  \sum_{\lambda}
  | \lambda + \tr (  \lambda \theta_{i} ) \,  \theta_{j}  \rangle
  \langle \lambda | \, .
\end{equation}
In particular, the curve $\beta = f ( \alpha) $ is transformed into 
\begin{equation}
\label{eq:XOR T}
  \beta  \mapsto    f ( \alpha ) + 
  \tr ( \alpha \theta_{j} ) f ( \theta_{i} ) +
  \tr [ f ( \alpha ) \theta_{i} ] \, \theta_{j}  + \tr ( \alpha \theta_{j} ) \,  
  \tr [ f ( \theta_{i} ) \theta_{i} ] \theta_{j} \, .
\end{equation}
Equivalently, the expansion coefficients are $\phi_{k} \mapsto
\phi_{k}+ \theta_{j}^{2^{k}} f ( \theta_{i} ) + f^{2^{k}} (
\theta_{i}) \, \theta_{j}$ . For example, for two qubits, the curve
$\beta =\alpha $ is transformed, under the action $\XOR_{12}$, into
$\beta =\sigma^{2} \alpha $; where $\sigma $ is a primitive element of
$\Gal{2^{2}}$, solution of the irreducible polynomial $x^{2}+
+1=0$. For three qubits, the curve $ \beta =\alpha $ becomes, also
under the action $XOR_{12}$, $\beta =\prim \alpha +
\alpha^{2}+\alpha^{4}$, where $\prim $ is the primitive element for
$\Gal{2^{3}} $, solution of $x^{3} + x +1=0$. In Table~\ref{tab:table}
we summarize the relevant information for the Galois fields used in
this paper.

%%%%%%%%%%%%%%%%%%%%%%%%%%%%%%%%%%%%%%%%
\begin{table}[t]
\caption{Irreducible polynomials and self-dual bases for the Galois
  fields  $\Gal{2^{n}}$ used in this paper.}
\label{tab:table}
\begin{indented}
\item[ ]
    \begin{tabular}{ccc}
\br
 Field & Irreducible polynomial & Self-dual basis  \\
      \mr
      $\Gal{2^{2}}$ & $x^{2}+x+1=0$ & $\{ \prim , \prim^{2} \} $ \\
      $\Gal{2^{3}}$ & $x^{3}+ x +1=0$ & 
      $ \{ \prim^{3}, \prim^{5}, \prim^{6} \}$ \\ 
      $\Gal{2^{4}}$ & $x^{4} + x +1=0$ & 
      $\{ \prim^{3},\prim^{7},\prim^{12},\prim^{13} \}$ \\
      $\Gal{2^{5}}$ & $x^{5} + x^{2} + 1 = 0 $ &
      $\{ \prim^{3}, \prim^{7}, \prim^{11}, \prim^{15}, \prim^{19} \}
      $ \\ 
\br
    \end{tabular}
  \end{indented}
\end{table}
%%%%%%%%%%%%%%%%%%%%%%%%%%%%%%%%%%%%%%%%

The $\SWAP$ operator exchanges the states of the $i$th and $j$th
qubits; i.e.,
\begin{equation}
  \SWAP_{ij}|\ldots ,a_{i},\ldots ,a_{j},\ldots \rangle
  =|\ldots ,a_{j},\ldots ,a_{i},\ldots \rangle \,.
\end{equation}
At the level of field elements, its action is 
\begin{equation}
  \alpha \mapsto \alpha +\varepsilon \tr(\alpha \varepsilon )\,,
  \qquad 
  \beta \mapsto \beta +\varepsilon \tr(\beta \varepsilon ) \,,
\end{equation}
where $\varepsilon =\theta _{i}+\theta _{j}$. Consequently, we can
write
\begin{equation}
\SWAP_{ij} = \sum_{\lambda }|\lambda +\varepsilon 
 \tr (\varepsilon \lambda )\rangle \langle \lambda |\,,
\end{equation}
so it transforms the curve $\beta =f(\alpha )$ into
\begin{equation}
  \beta \mapsto f(\alpha )+\tr(\alpha \varepsilon )f(\varepsilon )
  +\tr [ f ( \alpha ) \varepsilon ]\,\varepsilon +\tr(\alpha
  \varepsilon )
  \tr [ f ( \varepsilon ) \varepsilon ]\,\varepsilon \,.
\end{equation}
The expansion coefficients of the transformed curve are $\phi_{k}
+\varepsilon ^{2^{k}}f(\varepsilon )+f^{2^{k}}(\varepsilon)\,
\varepsilon $.

As our final nonlocal operation, we take squeezing, represented by
\begin{equation}
  \label{eq:squeez}
  S_{\xi} = \sum_{\lambda} | \xi \lambda \rangle \langle \lambda | \, ,
\end{equation}
whose action on the basic monomials is $S_{\xi} \, Z_{\alpha}
X_{\beta} \, S_{\xi}^{-1} = Z_{\xi\alpha} X_{\xi^{-1} \beta}$, so
that, the curve is transformed into $\beta = \xi^{-2} f(\alpha)$,
which, as expected, is just a global scaling

One can also define curve composition in such a way that given two
curves $ \beta =f (\alpha )$ and $\beta =g (\alpha )$, the function
$\beta =f ( g (\alpha ))$ is also a curve iff $f( g (\alpha )) = g 
(f (\alpha ))$.

The last remark concerning these curves is that they can always
transformed into the horizontal axis $\beta = 0$. Indeed, one can
always find an operator $P_{f}$ such that
\begin{equation}
  P_{f} Z_{\alpha} P_{f}^{-1} = \varphi_{f} (\alpha ) \, Z_{\alpha}
  X_{f(\alpha)} \, ,  
\label{P}
\end{equation}
where $\varphi_{f} (\alpha )$ is a phase factor. The operator $P_{f}$
can be conveniently expanded in the the basis $\{ |\widetilde{\lambda}
\rangle \}$ of eigenstates of $X_{\lambda}$
\begin{equation}
  P_{f} = \sum_{\lambda} c_{\lambda}^{(f)} \, |\widetilde{\lambda}\rangle
  \langle \widetilde{\lambda}|\, .
\end{equation}
The coefficients$c_{\lambda}^{(f)}$ can be determined by noting that
\begin{equation}
  P_{f} Z_{\alpha} P_{f}^{-1} = \sum_{\lambda} c_{\lambda +\alpha}^{(f)}
  c_{\lambda}^{(f) \ast} | \widetilde{\lambda +\alpha} \rangle 
  \langle  \widetilde{\lambda} | \, ,
\end{equation}
while 
\begin{equation}
  Z_{\alpha} X_{f(\alpha)} =\sum_{\lambda} \chi [ \lambda f ( \alpha )] | 
  \widetilde{\lambda +\alpha} \rangle \langle \widetilde{\lambda} | \, ,
\end{equation}
so that $c_{\lambda}^{(f)}$ satisfy the recurrence relation 
\begin{equation}
  c_{\lambda +\alpha}^{(f)} \, c_{\lambda}^{(f) \ast} = \phi_{f} ( \alpha ) \,
  \chi [ \lambda f ( \alpha ) ] \, .
\end{equation}
The phase $\phi_{f} ( \alpha ) $ is evaluated by substituting $\lambda
=\alpha $ in the above equation and taking $c_{0}^{(f)}=1$, which
finally leads to
\begin{equation}
  \label{eq:recrel}
  c^{(f)}_{\lambda} \, c^{(f)}_{\lambda^{\prime}} = 
  \chi [ \lambda^{\prime} f(\lambda) ] \, 
  c^{(f)}_{\lambda+\lambda^{\prime}} \, .
\end{equation}
The solution of equation~(\ref{eq:recrel}) is not unique, but choosing
all $ c_{\theta_{i}}^{(f)}$ for a basis to be positive, we arrive at
the following explicit expression
\begin{equation}
  c_{\lambda }^{( f) } = \left [ \prod_{i=1}^{n} 
    \sqrt{ (-1)^{\ell_{i} M_{ii}^{(f)}}} \right ] 
  (-1)^{\bm{\ell}^{t} \tilde{\mathbf{M}} \bm{\ell} } \, .
\end{equation}
Here $\bm{\ell}^{t}=(\ell_{1},\ldots ,\ell_{n})$ is the vector of the
expansion coefficients of $|\lambda \rangle $ in the self-dual basis,
and the principal branch of the square root is chosen. In addition, $
M_{ij}^{(f)}=\tr [ \prim_{i} f ( \prim_{j} ) )] $ and
\begin{equation}
  \widetilde{\mathbf{M}} = \left \{ 
    \begin{array}{ll}
      M_{ij}^{(f)} & \quad j < i \, , \\ 
      &  \\ 
      0 & \quad j > i \, .
    \end{array}
  \right.  \label{M}
\end{equation}
All this means that given a curve we can immediately obtain the
eigenstates $ |\lambda ; f\rangle $ of the set of commuting monomials
corresponding to this curve: $|\lambda ; f \rangle = P_{f} |\lambda
\rangle $, with $| \lambda \rangle $ being the computational basis.

In a similar way, the curves $\alpha =g(\beta )$ can be transformed
into the vertical axis $\alpha =0$:
\begin{equation}
  Z_{g (\beta)} X_{\beta} = \psi ( \beta ) \, Q_{g} X_{\beta} Q_{g}^{-1} \, ,
  \label{Q}
\end{equation}
where $\psi (\beta) $ is a phase. Again, one can show that $Q_{g}$ can
be expressed as
\begin{equation}
  Q_{g} = \sum_{\kappa} c^{(g)}_{\kappa} \, |\widetilde{\kappa} \rangle
  \langle \widetilde{\kappa} | \, ,
\end{equation}
and the coefficients $c_{\kappa}^{\left( g\right)}$ satisfy
(\ref{eq:recrel} ).

\section{Graph states and their phase space representation}

Consider a graph $G=\{ V, E \}$, which is a set of vertices $V = \{ 1,
\ldots, n \}$ connected in a specific way by a set of edges $E \subset
V \times V$, specifying the neighborhood relation between
vertices~\cite{Diestel:2000fj}. In the following, we only consider
simple graphs, containing neither loops neither multiple edges. To
each vertex $i$ we attach a single qubit and each edge $\{i, j\}$
represents the interaction between the corresponding qubits. With the
graph we associate a complete set of $n$ commuting monomials
\begin{equation}
  \label{monomials}
  K_{i} = \sigma_{z}^{(i)} \, \prod_{j \in N_{i}} \sigma_{x}^{(j)} \, ,
\end{equation}
where $N_{i}$ is the neighborhood for that vertex, $N_{i}= \{ j \in V
| \{i, j\} \in E \}$ (i.e., all the vertices connected to
$i$). Please, note that we have interchanged the order of $\sigma_{z}$
and $\sigma_{x}$ with respect to the standard
definition~\cite{Hein:2006fk}: this corresponds to a global Clifford
$y$-rotation, and will allow us to work with curves $\beta =
f(\alpha)$ and not $\alpha = g(\beta)$.

The associated graph state $| G \rangle$ is just the unique common
eigenvector of $\{ K_{i} \}$ with all eigenvalues +1. The monomials
$\{ K_{i} \}$ are also the generators of the stabilizer of $| G
\rangle$~\cite {Gottesman:1997oq}, which is an Abelian subgroup of the
local Pauli group $ \mathcal{P}_{n}$~\cite{Chuang:2000fk}. Any
stabilizer state is locally equivalent to some graph
state~\cite{Van-den-Nest:2004fk} (more exactly, they are equivalent
under local Clifford transformations).

The crucial observation for what follows is that the graph-state
generators $ \{K_{i}\}$ can be written precisely as $\{Z_{\alpha}
X_{\beta =f(\alpha )}\}$ , when $\alpha $ runs the self-dual basis
$\{\theta_{1},\ldots ,\theta _{n}\} $, if the function $\beta =f
(\alpha )$ satisfies the extra condition
\begin{equation}
  \tr [ \theta_{i} \, f (\theta_{i}) ]=0 \,. 
  \label{f_gs}
\end{equation}
Note in passing that the $x$-rotation $U_{x}^{\xi}$, with
\begin{equation}
  \xi =\sum_{i} \tr [ f ( \theta_{i}) \theta_{i} ] \, \theta_{i} \, ,
  \label{LT1}
\end{equation}
reduces any curve $\beta =f(\alpha )$ to the form $\beta = f(\alpha)
+\sum_{i} \tr[ f (\theta_{i}) \theta_{i} ]\,\tr(\alpha
\theta_{i})\,\theta_{i}$, which obviously satisfies the graph-state
condition~(\ref{f_gs}). For instance, in $\Gal{2^{3}}$, the non-graph
curve $\beta =\prim^{4} \alpha $ is transformed into the graph curve
$\beta =\prim^{4}\alpha^{2} +\prim^{2}\alpha^{4}$ by two $x$-rotations
in qubits corresponding to elements of the self-dual basis $\prim^{3}$
and $\prim^{5} $: $u_{x}^{\prim^{3}}u_{x}^{\prim^{5}}$.

It is indeed possible to find a generic form of the curves that
fulfill (\ref{f_gs}). To this end, let us substitute the explicit
expansion (\ref{RC}) into the condition (\ref{f_gs}):
\begin{equation}
  \sum_{i=0}^{n-1}\tr(\phi _{i}\theta _{k}^{2^{i}}\theta _{k})=0\,,
 \quad
  k=1,\ldots ,n.  
  \label{odd n}
\end{equation}
Since the expansion coefficients $\phi _{i}$ can be expressed in the
form of equation~(\ref{phi_f}), the condition $\tr(\phi _{0})=0$
should be always satisfied for graph-state curves. When $n$ is odd,
and by using the property (\ref{cc}), we can reduce the sum (\ref{odd
  n}) to the simpler form
\begin{equation}
  \tr(\phi _{0}^{2^{n-1}}\theta _{k})=0 \, ,
\end{equation}
which shows that any curve for an odd number of qubits with 
$\phi_{0}=0$ corresponds to a graph.

When $n$ is even, we have to take into account an additional condition
on the coefficients, $\phi_{n/2}=\phi_{n/2}^{2^{n/2}}$, which leads to
the restriction
\begin{equation}
  \phi_{0}=\sum_{i=1}^{n} \tr(\phi_{n/2} \theta_{i}^{2^{n/2}} \theta_{i}) \,
  \theta_{i}^{2}\, .
\end{equation}

A graph can be also characterized by its adjacency matrix $\Gamma $,
which is a symmetric $n\times n$ matrix such that
\begin{equation}
  \Gamma _{ij}=\left\{ 
    \begin{array}{ll}
      1, & \quad \{i,j\}\in E\, \\ 
      &  \\ 
      0, & \quad \{i,j\}\notin E\,.
    \end{array}
  \right. 
\end{equation}
The adjacency matrix for graphs associated to the curves fulfilling
(\ref {f_gs}) has the form
\begin{equation}
  \Gamma _{ij}=\tr [ \theta_{i} f(\theta_{j}) ] \, . 
  \label{G}
\end{equation}
For an arbitrary curve $\beta =f(\alpha )$, the result still holds
once the diagonal part of $\Gamma _{ij}$ in (\ref{G}) is
removed. Note, however, that the same adjacency matrix may correspond
to different curves. For instance, for three qubits the curves on
$\Gal{2^{3}}$ $\beta =\prim^{4}\alpha $ and $\beta =\prim^{4}
\alpha^{2}  +\prim^{2} \alpha ^{4}$ have the same adjacency
matrix: the second curve corresponds to a graph, while the first one
does not.

Given the matrix $\Gamma $ of a graph $G$, one can use (\ref{G}) to
immediately find that the corresponding $f(\theta_{j})$ are
\begin{equation}
  f(\theta_{j}) = \sum_{i=1}^{n} \tr [ \theta_{i} f ( \theta_{j} )] \,
  \theta_{i} = \sum_{i=1}^{n}\Gamma_{ij} \,\theta_{i} \, .  
\label{f2}
\end{equation}
Multiplying (\ref{f2}) by $\alpha_{j} = \tr (\alpha \theta_{j})$ and
summing we obtain
\begin{equation}
  \sum_{j=1}^{n} \alpha_{j} f( \theta_{j} ) = 
 \sum_{j=1}^{n} f  (\alpha_{j}\theta_{j} ) = 
 f(\alpha ) = \sum_{i,j=1}^{n} \alpha_{j}  
\Gamma_{ij} \theta_{i} \,,  
\label{_03}
\end{equation}
so the curve corresponding to the matrix $\Gamma $ is
\begin{equation}
  \beta =\sum_{i,j=1}^{n} \tr(\alpha \theta_{j}) \Gamma_{ij}\theta_{i} =
  \sum_{k=1}^{n}\sum_{i,j=1}^{n} \Gamma_{ij}\theta_{j}^{2^{k}}\theta_{i}
  \alpha^{2^{k}} \,,
\end{equation}
or, equivalently, its expansion coefficients $\phi_{k}$ are
\begin{equation}
  \phi_{k}=\sum_{i,j=1}^{n} \Gamma_{ij} \,\theta_{i}^{2^{k}}\theta_{j} \, .
\end{equation}

Let us work out some examples. The simplest one is perhaps a linear chain of 
$n$ qubits, with adjacency matrix 
\begin{equation}
  \Gamma =\left( 
    \begin{array}{ccccccc}
      0 & 1 & 0 & 0 & 0 & 0 & 0 \\ 
      1 & 0 & 1 & \ddots  & \ddots  & 0 & 0 \\ 
      0 & 1 & 0 & 1 & 0 & \ddots  & 0 \\ 
      \vdots  & 0 & \ddots  & \ddots  & \ddots  & 0 & \vdots  \\ 
      0 & \vdots  & 0 & 1 & 0 & 1 & 0 \\ 
      0 & 0 & 0 & 0 & 1 & 0 & 0
    \end{array}
  \right) \,,
\end{equation}
so the associated curve is
\begin{equation}
\beta = \sum_{i=2}^{n-1} \tr(\alpha \theta _{i})
 (\theta _{i+1}+\theta_{i-1}) +
\tr(\alpha \theta _{1})\,\theta _{2} +
\tr(\alpha \theta _{n})\,\theta_{n-1}\,.  
\label{eq:flin}
\end{equation}
Another interesting example is a star graph, in which every single
vertex is connected only to the center vertex (which we take as the
$n$th qubit). Now the adjacency matrix is
\begin{equation}
  \Gamma =\left( 
    \begin{array}{ccccc}
      0 & 0 & ... & 0 & 1 \\ 
      0 & \ddots  & \ddots  & 0 & 1 \\ 
      \vdots  & \ddots  & \ddots  & \vdots  & \vdots  \\ 
      0 & 0 & \cdots & 0 & 1 \\ 
      1 & 1 & \ddots  & 1 & 0
    \end{array}
  \right) \,,  
  \label{_b001}
\end{equation}
which leads to the very simple curve
\begin{equation}
  \beta =\,\theta _{n}+\tr(\alpha \theta _{n})\,.  
  \label{_b008}
\end{equation}
Finally, we consider a graph consisting of a closed polygon, with
matrix
\begin{equation}
  \Gamma =\left( 
    \begin{array}{cccccc}
      0 & 1 & 0 & \cdots  & 0 & 1 \\ 
      1 & 0 & 1 & 0 & \ddots  & 0 \\ 
      0 & 1 & \ddots  & \ddots  & \ddots  & \vdots  \\ 
      \vdots  & \ddots  & \ddots  & 0 & 1 & 0 \\ 
      0 & \ddots  & 0 & 1 & 0 & 1 \\ 
      1 & 0 & \cdots  & 0 & 1 & 0
    \end{array}
  \right) \,,
\end{equation}
and associated curve
\begin{equation}
  \beta = \sum_{i=1}^{n} \tr(\alpha \theta _{i+1})
    (\theta _{i+2}+\theta _{i})\,,
\end{equation}
where the indices of the basis elements are taken $\bmod n$.

Please, observe carefully that for each of the examples worked out so
far the following relation holds:
\begin{equation}
  \beta (\theta _{i}) =\sum_{j\in N_{i}} \theta _{j} \, ,
\end{equation}
that is, the curve evaluated at each vertex appears as a sum of basis
elements associated with all the connected vertices.  In figure~1 we
graphically summarize these relevant examples of the correspondence
between phase-space curves and graphs.

%%%%%%%%%%%%%%%%%%%%%%%%%%%%%%%%%%%%%%%%%%%%%
\begin{figure}[tbp]
\centering
\includegraphics[height=6cm]{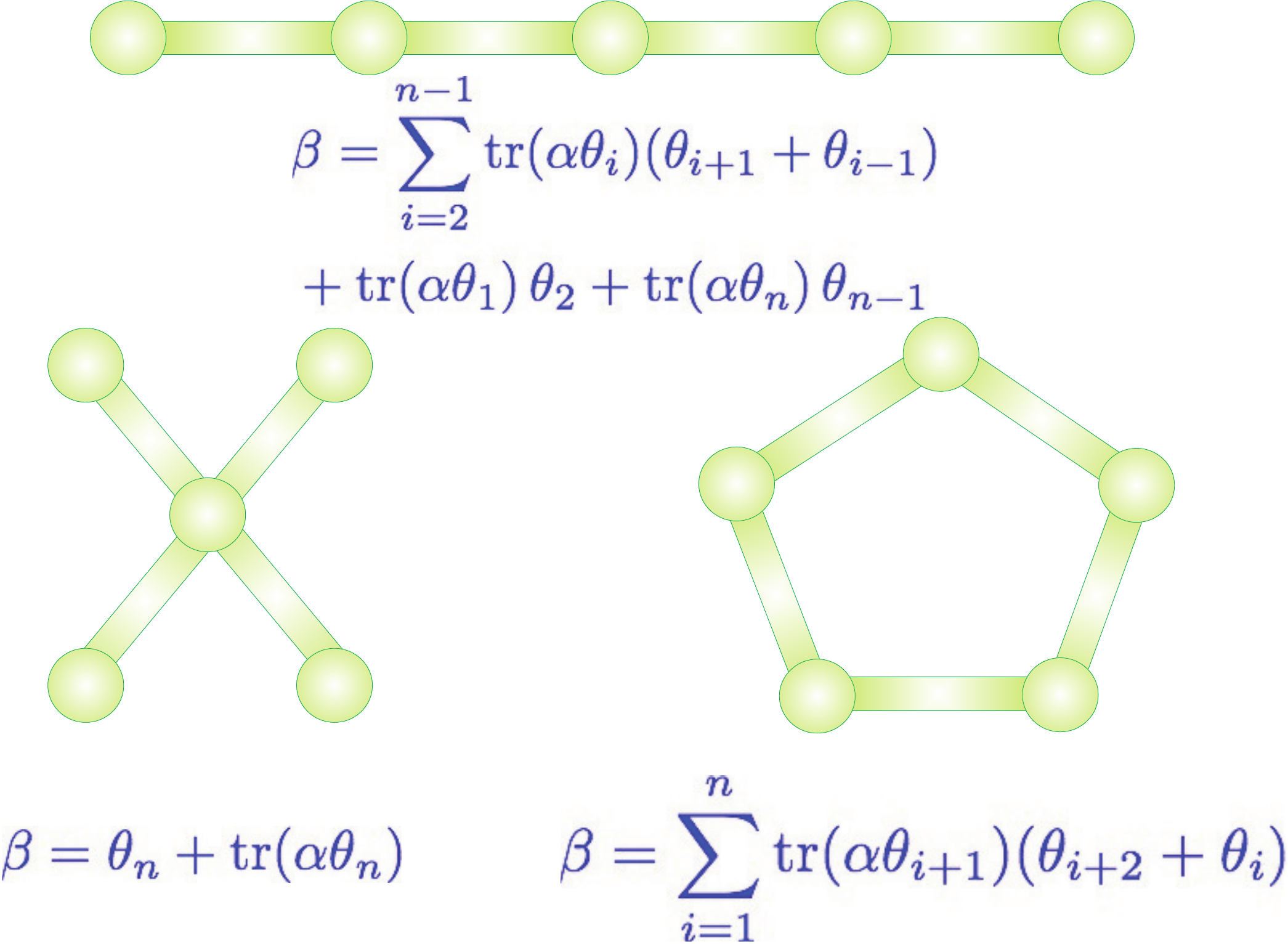}
\caption{(Color online) Curves associated with different graph states. In
all the cases, we have restricted ourselves to five qubits, although the
curve equations are written for an arbitrary number $n$.}
\end{figure}
%%%%%%%%%%%%%%%%%%%%%%%%%%%%%%%%%%%%%%%%%%%%

Concerning the nonlocal operations $\XOR$ and $ \SWAP$, one can check
that graph curves are transformed into graph curves.

Local Clifford operations on graph states can be easily explained in terms
of the local complementation. For the graph $G = \{ V , E\}$, the local
complement of $G$ at $i$ is obtained by complementing the subgraph of $G$
induced by the neighborhood $N_{i}$ of $i$ and leaving the rest of the graph
unchanged. This is implemented by the local unitary 
\begin{equation}
V_{i} = u_{x}^{i} \prod_{j\in N_{i}} u_{z}^{j}
\end{equation}
The curve coefficients transform thus in a simple way 
\begin{equation}
\phi_{k} \underset{V_{i}}{\mapsto} 
\phi_{k}+ f^{2^{k}+1} ( \theta_{i} ) + \sum_{i=1}^n \tr [ f (
\theta_{i} ) \theta_{i} ] \, 
\theta_{i}^{2^{k} +1} \,.
\end{equation}
It is clear form that the local complement does not change the
coefficient $ \phi_{0}$, which reflects the fact that a graph is also
transformed into a graph.

Finally, it is worth noting that for graph curves, the coefficients $
c_{\lambda }^{(f)}$ in the expansion (\ref{eq:recrel}) simplify to
\begin{equation}
  c_{\lambda }^{(f)}=(-1)^{\frac{1}{2}\bm{\ell}^{t} \Gamma \bm{\ell}}\, ,
\end{equation}
where $\Gamma$ is the adjacency matrix corresponding to the curve.

\section{Factorization structure}

One essential property of these curves is their factorization
structure, which is related to the commutation condition between
blocks of single-qubit operators.

Consider the commuting monomials labeled by points of a given
curve. Each monomial is a direct product of $n$ Pauli operators. Let
us divide each monomial into two parts, so that the first part
contains $k$ Pauli operators, corresponding to the $(i_{1}, \ldots,
i_{k})$ qubits and the second part $n-k$ operators of the rest of the
qubits. If any Pauli operator in the first block commutes with all the
others in that block, we will say that the corresponding set of
monomials is factorized at least into two subsets. Obviously, the
second blocks would then also commute among themselves. Moreover,
inside the first or second blocks some sub-blocks may exist that
commute with corresponding sub-blocks, etc. Thus, we can represent any
curve $\gamma \in \Gal{2^{n}}$ in the following factorized form:
\begin{equation}
  \gamma =\{m_{1},m_{2},\ldots ,m_{N}\} \, ,  
\label{1_curve_part}
\end{equation}
where $\quad 0<m_{1} \leq m_{2} \leq \ldots \leq m_{N}$ and $m_{i}\in
\mathbb{N}$ is the number of particles in the $i$th block that cannot
be factorized anymore. It is clear that $\{m_{1},m_{2},\ldots,
m_{N}\}$ is just a partition of the integer $n$, so the maximum number
of terms is $n$, which corresponds to a completely factorized curve,
$\gamma = \underbrace{ \{1,1,\ldots ,1\}}_{n}$, and the minimum number
of terms is one, corresponding to a completely nonfactorized curve
$\gamma =\{n\}$.

The rays (straight lines passing through the origin) $\alpha =0$, $\beta =0,$
and $\beta =\alpha $ are always completely factorized, while all the other
are apparently completely nonfactorized for any dimension, except the
special case of four qubits, where there are two locally equivalent rays
with factorization (2,2).

It is easy to convince oneself that completely factorized curves can
be written as
\begin{equation}
  f(\alpha )=\sum_{i=1}^{n} \tr (\alpha \theta_{i}) \,\xi_{i}\,\theta_{i} \, ,
\end{equation}
where $\xi_{i}$ are the components of an arbitrary $\xi \in
\Gal{2^{n}}$ in the self-dual basis.

For bifactorized curves, whose adjacency matrix is just a $2k \times
2k$ permutation matrix, the corresponding form is
\begin{equation}
f ( \alpha ) = \sum_{n =1}^{k} [ \tr ( \alpha \theta_{i_{n}} ) \,
\theta_{j_{n}} + \tr ( \alpha \theta_{j_{n}} ) \, \theta_{i_{n}} ] \, ,
\label{_H002}
\end{equation}
where $\{ i_{1}, j_{1} \} , \{ i_{2}, j_{2} \}, \ldots , \{ i_{k}, j_{k} \} $
are ordered pairs of connected vertices. In this case, two relations are
automatically satisfied: $f ( f ( \alpha ) ) = \alpha$ and $\tr [ 
f ( \alpha ) ] =\tr ( \alpha ) $.

In general the problem of fully determining the factorization of a
curve is quite difficult, and basically corresponds to the separation
of an adjacency matrix in independent blocks by permutation of lines
and columns. The number of such blocks can be determined by, e.g.,
counting the number of zero eigenvalues of the Laplacian
matrix~\cite{Beineke:2004qy}. Nevertheless, some general
considerations can be made. Let us evaluate a function $ f(\alpha )$,
corresponding to a graph, in an arbitrary element $\theta_{i}$ of a
self-dual basis,
\begin{equation}
  f(\theta_{i}) = \sum_{j=1}^{n}\tr [ f ( \theta_{i} ) \theta_{j} ]\,
  \theta_{j} =\sum_{j=1}^{n} \Gamma_{ij}\,\theta_{j}\,,
\end{equation}
which is the sum of elements of the basis corresponding to vertices
(qubits) connected to the $i$th vertex. This means that this $i$th
vertex is disconnected form the rest of the graph if
$f(\theta_{i})=0$. Moreover, for each disconnected subgraph $A_{i}$,
the function $f(\alpha )$ maps the subgraph onto itself
$f(A_{i})=A_{i}$ and thus
\begin{equation}
  \tr f \left( \sum_{j\in A_{i}} \theta_{j} \right) = 0
\end{equation}
for each separated block.

%%%%%%%%%%%%%%%%%%%%%%%%%%%%%%%%%%%%%%%%%%%%%%%%%%%%%%
\begin{figure}[tbp]
\centering
\includegraphics[height=7cm]{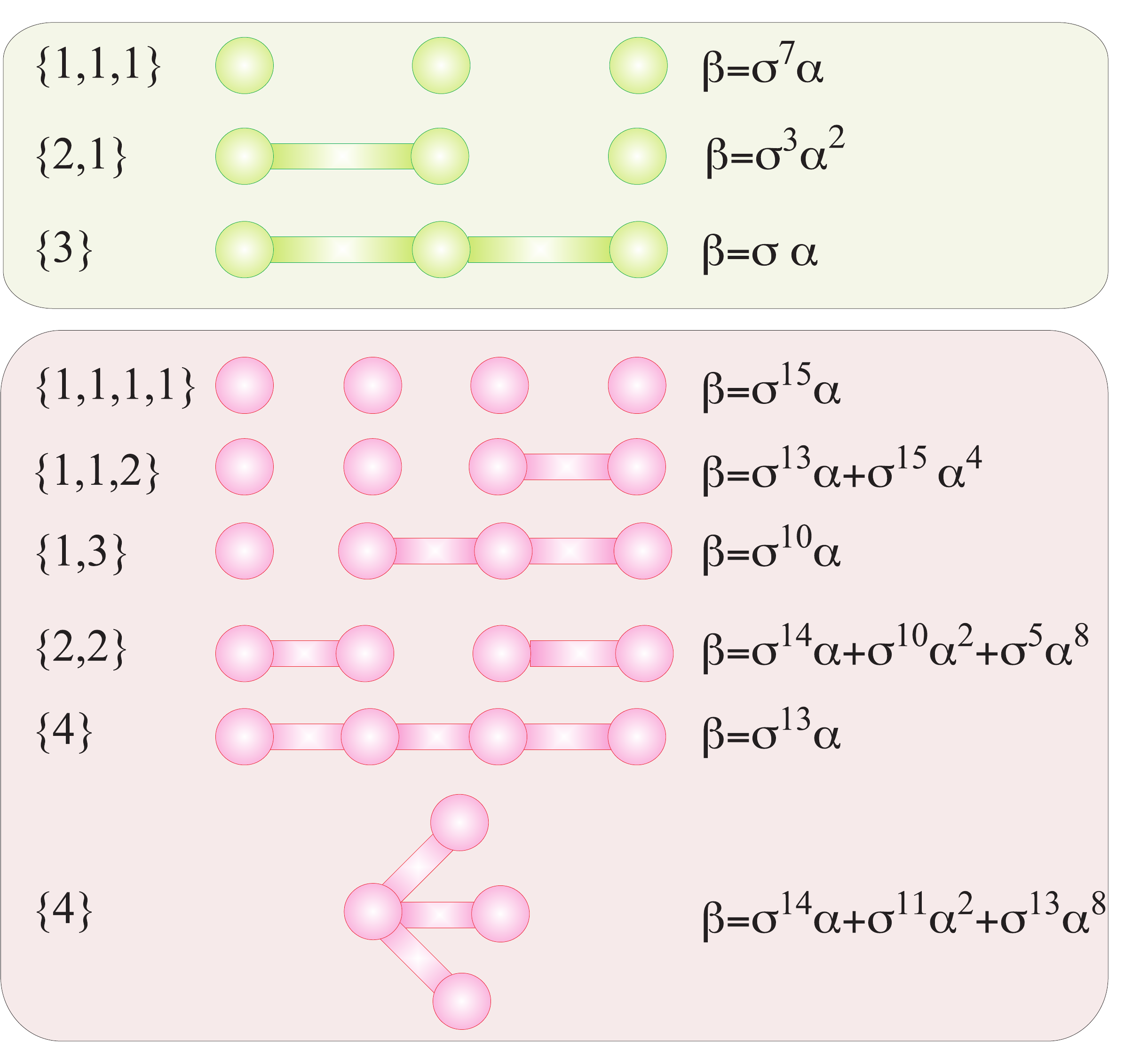}
\caption{(Color online) Inequivalent curves for three (top) and four
(bottom) qubits. In the left, we show the factorization structure; in the
right, the curve in the simplest algebraic form and in the middle the
locally equivalent graphs.}
\end{figure}
%%%%%%%%%%%%%%%%%%%%%%%%%%%%%%%%%%%%%%%%%%%%%%%%%%%%%%

In figure~2 we list the locally inequivalent curves for three and four
qubits, where they are expressed in the simplest algebraic form (not
necessarily corresponding to graphs) and the equivalent graphs are
plotted.  This can be easily continued to higher number of qubits.

\section{Some applications}

\subsection{Local and nonlocal transformations}

The algebraic properties of the commuting stabilizers $\{Z_{\alpha}
X_{f( \alpha)} \}$ allows us to scrutinize in a very simple way the
action of both local and nonlocal transformations. Moreover, given
then relation between curves and adjacency matrices we can easily find
how graphs are transformed, avoiding in this way involved
procedures~\cite{Van-den-Nest:2004fk}.

Let us start by considering the linear operations treated in
section~3. It is clear from (\ref{x}) that $x$-rotations (labelled by
the parameter $\xi$, whose nonzero components indicate the qubits
subjected to rotations) do not change the graph; they add only a
diagonal element to the adjacency matrix: $ \Gamma \mapsto \Gamma +
\mathrm{diag} ( \xi_{1}, \ldots,\xi_{n})$.

On the other hand, $z$-rotations change the adjacency matrix in a
nontrivial way, as hinted by (\ref{x}). First, we find the
corresponding curve for which we change the variable $\alpha \mapsto
\mu =\alpha + \sum_{i=1}^{n}\xi_{i} \tr [ f ( \alpha ) \theta_{i} ] \,
\theta_{i}$. This requires the inversion of the system
\begin{equation}
  \mu_{i}=\alpha_{i} +\xi_{i} \sum_{j=1}^{n} \alpha_{j} \, \Gamma_{ij} \, ,
  \label{z mu}
\end{equation}
to be able to determine $\alpha =\alpha ( \mu )$ and thus find the set
$ \{Z_{\mu} X_{g( \mu)} \}$, where $g= f ( \alpha ( \mu ) )
$. Finally, the curve $\beta =g( \mu )$ can be converted into a
graph-curve (if necessary) by some $x$-rotations, which immediately
gives the new adjacency matrix.

As an example, let us analyze a four-qubit graph with adjacency matrix
\begin{equation}
  \Gamma =\left ( 
    \begin{array}{cccc}
      0 & 1 & 0 & 0 \\ 
      1 & 0 & 1 & 0 \\ 
      0 & 1 & 0 & 1 \\ 
      0 & 0 & 1 & 0
    \end{array}
  \right ) \, ,  
  \label{G1}
\end{equation}
and find how it is transformed under $z$-rotations of the second and
the fourth qubits. The curve corresponding to the matrix (\ref{G1}) is
$\beta =\prim^{7} \alpha^{2} + \prim^{11} \alpha^{8}$. The rotation
$\openone \otimes u_{z} \otimes \openone \otimes u_{z}$ transforms it
into the curve $\beta = \sigma^{11} \alpha +\sigma^{5} \alpha^{2} +
\sigma^{15} \alpha^{4} + \sigma^{10} \alpha^{8}$. This means that
\begin{equation}
  \Gamma \mapsto \Gamma^{\prime} = \left ( 
    \begin{array}{cccc}
      0 & 1 & 0 & 0 \\ 
      1 & 0 & 1 & 1 \\ 
      0 & 1 & 0 & 1 \\ 
      0 & 1 & 1 & 1
    \end{array}
  \right ) \, .
\end{equation}
The final adjacency matrix is obtained from the above matrix just by
removing the diagonal elements (which can be done just by applying the
$x$-rotation $\openone \otimes \openone \otimes \openone \otimes
u_{x}$). The resulting graph-curve is $\beta =\prim^{6} \alpha^{2} +
\prim^{10} \alpha^{4} + \prim^{3}\alpha^{8}$.

Next, we turn our attention to nonlocal operations. According to (\ref
{eq:XOR T}) the adjacency matrix associated with the graph-curve
transformed by $\XOR_{ij}$ is
\begin{equation}
  \Gamma_{ij} \mapsto \Gamma_{kl}^{^\prime} = 
  \tr [ \theta_{k} f ( \theta_{l})] + 
  \tr [ \theta_{k} f ( \theta_{i} )] \delta_{lj} + 
  \tr [ \theta_{i} f ( \theta_{l} ) ] \delta_{kj} \, .  
  \label{_0002}
\end{equation}
This can be rewritten as $\Gamma \mapsto \Gamma^{\prime} = G \, \Gamma
\, G^{-1}$, where $G_{kl }= \delta_{kl} + \delta_{kj} \delta_{li}$ is
a triangular matrix with $\det \, G=1$. For instance, the application
of the gate $\XOR_{12}$ to the graph defined by the matrix (\ref{G1} )
leads to
\begin{equation}
  \Gamma^{\prime} = \left ( 
    \begin{array}{cccc}
      0 & 1 & 1 & 0 \\ 
      1 & 0 & 0 & 1 \\ 
      1 & 0 & 0 & 0 \\ 
      0 & 1 & 0 & 0
    \end{array}
  \right ) \, ,
\end{equation}
and the corresponding curve is $\beta =\prim^{14} \alpha^{2} +
\prim^{7} \alpha^{8}$.

Finally, the squeezing operation (\ref{eq:squeez}) transforms the
curve $ \beta =f ( \alpha ) $ into $\beta =f ( \mu \xi^{-1} )
\xi^{-1}$ and, thus the adjacency matrix $\Gamma $ corresponding to
the initial graph into
\begin{eqnarray}
  \Gamma_{ij} & \mapsto & \tr [ \theta_{i} f ( \theta_{j} \xi^{-1} ) \xi^{-1} ]
  = \tr \left [ \sum_{k=1}^{n} \tr ( \theta_{i} \xi^{-1} \theta_{k} )
    \theta_{k} \; f\left( \sum_{l=1}^{n} \tr ( \theta_{j} \xi^{-1} \theta_{l} )
      \theta_{l} \right) \right ]  
\nonumber \\
  & = & \sum_{k,l=1}^{n} \tr ( \theta_{i} \xi^{-1} \theta_{k} )
 \tr [   \theta_{k} f ( \theta_{l} ) ] \tr ( \theta_{j} \xi^{-1}
 \theta_{l} ) =
M_{ik}   \Gamma_{kl} M_{lj} \, ,
\end{eqnarray}
where $M_{ik} = \tr ( \theta_{i} \xi^{-1} \theta_{k}) $ is the matrix
$M$ for the ray $\beta=\xi^{-1}\alpha $.

For example, the graph corresponding to the adjacency matrix
(\ref{G1}) is transformed under $S_{\prim}$ (where $\prim $ is the
primitive element for $\Gal{2^{4}}$) converts into the graph defined
by
\begin{equation}
  \Gamma^{\prime}=
\left ( 
    \begin{array}{cccc}
      0 & 1 & 1 & 1 \\ 
      1 & 0 & 0 & 1 \\ 
      1 & 0 & 0 & 0 \\ 
      1 & 1 & 0 & 0
    \end{array}
  \right ) \, ,
\end{equation}
and the corresponding curve is $\beta =\prim^{4}\alpha^{2} +
\prim^{2} \alpha^{8}$.

\subsection{Computing reduced density matrices}

As final illustration of the potential of our method, we consider the
density matrix of the graph state corresponding to the curve $\beta =
f( \alpha )$, which is given by
\begin{equation}
  \varrho = 2^{-n} \sum_{\alpha} Z_{\alpha} X_{f ( \alpha)} \, .
\end{equation}
Now assume that we trace over $n-k$ qubits. This leads to the
following reduced density matrix
\begin{equation}
  \tilde{\varrho} =2^{-k} 
  \sum_{\alpha_{1},\ldots ,\alpha_{k} \in \mathbb{Z} _{2}} 
  Z_{\alpha_{1} \theta_{1}+\ldots +\alpha_{k}\theta_{k}} 
  X_{f ( \alpha_{1}\theta_{1}+ \ldots +\alpha_{k}\theta_{k})} \, .
\end{equation}
With the machinery developed so far, this can be recast in the
explicit form
\begin{equation}
  \fl \tilde{\varrho} =2^{-k} 
  \sum_{\alpha_{1},\ldots ,\alpha_{k} \in \mathbb{Z }_{2}} 
  \prod_{i=1}^{k} \otimes \sigma_{z}^{\alpha_{i}} 
  \sigma_{x}^{\tr [ f (
    \alpha_{1}\sigma_{1}+\ldots+\alpha_{k}\sigma_{k}) \sigma_{i}]} = 
  2^{-k}  \sum_{\alpha_{1},\ldots ,\alpha_{k} \in \mathbb{Z}_{2}} 
  \prod_{i=1}^{k} \otimes \sigma_{z}^{\alpha_{i}} 
  \sigma_{x}^{ \sum_{p=1}^{k}\alpha_{p}\Gamma_{pi}} \, ,
\end{equation}
where $\Gamma $ is the adjacency matrix of the initial graph.

\section{Conclusions}

In summary, we have developed a phase-space picture of stabilizer
states, i.e., a complete set of commuting monomials in the generalized
Pauli group.  This alternative description identifies these states
with functions (which can be associated with specific nonsingular
curves in discrete phase-space) over the finite field $\Gal{2}^{n}$.

All the operations on graph states can be expressed in this language
as transformations of the curves. Some of these transformations, such
as the $ \mbox{\textsc{xor}}$ gate or the discrete squeezing, are
simpler to apply to the algebraic functions than to the graphs. On the
contrary, the factorization is more involved at the algebraic
level. We expect that the proposed method can be used for
classification of graph states when the number of qubits is large and
graphs do not provide any useful information.  Work in this direction
is in progress.

\ack We would like to thank the two anonymous referees for their
constructive comments. This work is partially supported by the Grant
106525 of CONACyT (Mexico), the Grants FIS2008-04356 and FIS2011-26786
of the Spanish DGI and the UCM-BSCH program (Grant GR-920992).

\newpage

%\bibliographystyle{iopart-num}
%\bibliography{Graphs}

\providecommand{\newblock}{}

\end{document}